\documentclass[a4paper,10pt,twoside]{cpc-hepnp}

\usepackage{multicol}
\usepackage{graphicx}
\usepackage{booktabs}
\usepackage{amssymb,bm,mathrsfs,bbm,amscd}
\usepackage[tbtags]{amsmath}
\usepackage{lastpage}

\begin{document}

\fancyhead[co]{\footnotesize Sinya Aoki: Baryon-Baryon Interactions from Lattice QCD}

\footnotetext[0]{Received 31 December 2009}

\title{Baryon-Baryon Interactions from Lattice QCD
}

\author{%
      Sinya Aoki$^{1;1)}$\email{saoki@het.ph.tsukuba.ac.jp}%
      \quad for HAL QCD Collaboration}
\maketitle

\address{%
1~(Graduate School of Pure and Applied Sciences,  University of Tsukuba, Tsukuba, Ibaraki 305-8571, Japan)\\
}

\begin{abstract}
We report on new attempt to investigate baryon-baryon interactions in lattice QCD.
From the Bethe-Salpeter (BS) wave function, we have successfully extracted
the nucleon-nucleon ($NN$) potentials in quenched QCD simulations, which reproduce qualitative features of modern $NN$ potentials. The method has been extended to obtain the tensor potential as well as the central potential and also applied to the hyperon-nucleon ($YN$) interactions, in both quenched  and full QCD.
\end{abstract}



\begin{multicols}{2}

\section{Introduction}
 In 1935 Yukawa introduced virtual particles, pions, to explain  the nuclear force\cite{Yukawa},  which bounds protons and neutrons inside nuclei. Since then  the nucleon-nucleon ($NN$) interaction has been extensively  investigated at low energies both theoretically and experimentally.  Fig.~\ref{fig:potential} shows modern $NN$ potentials, which are characterized by the following features\cite{Taketani,Machleidt}.  
 \begin{center}
\includegraphics[width=75mm]{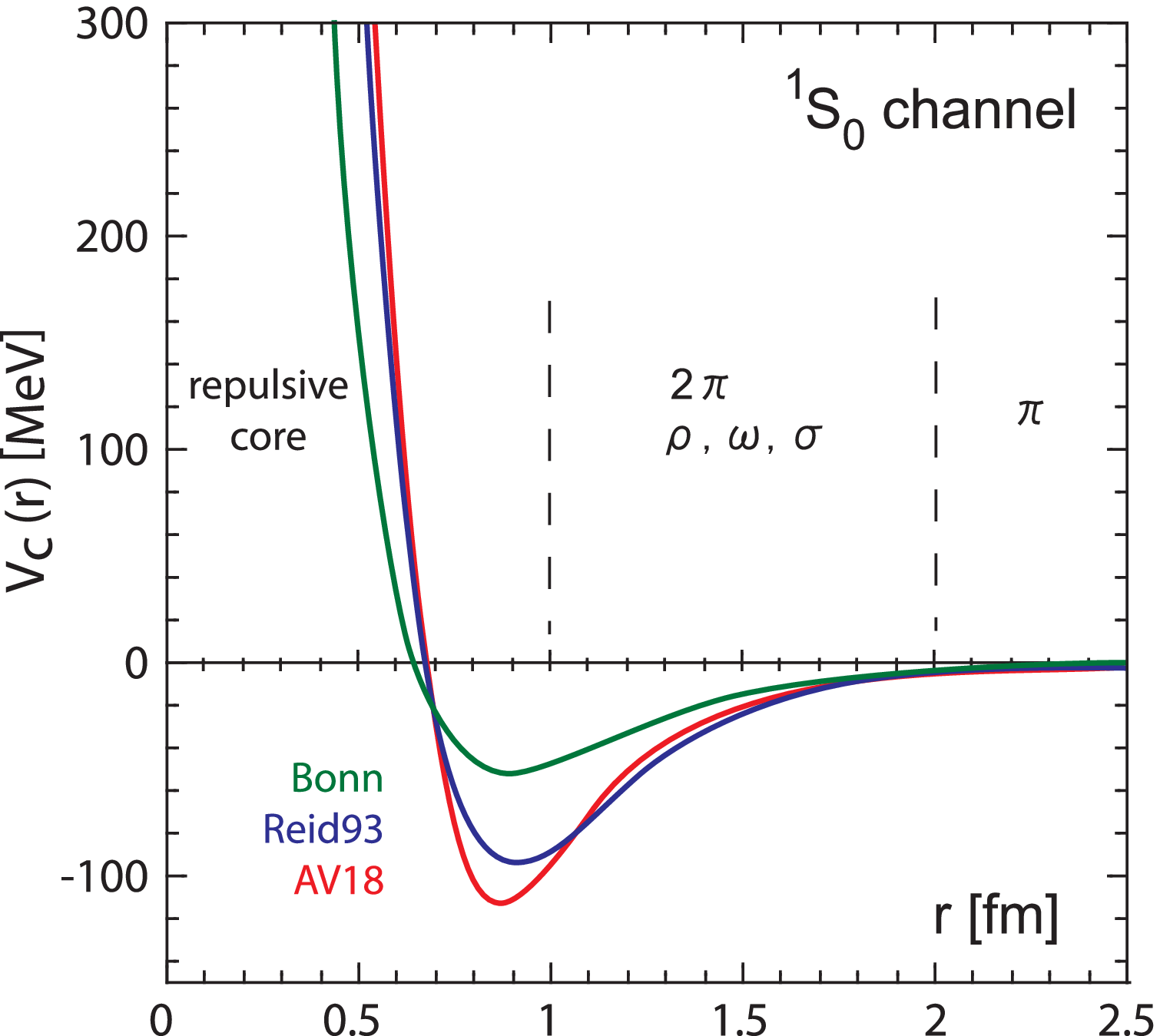}
\figcaption{\label{fig:potential}Three examples of the modern $NN$ potential for the $^1S_0$ (spin singlet and $S$-wave) state: Bonn\protect\cite{Bonn}, Reid93\protect\cite{Reid93} and AV18\protect\cite{AV18}.}
\end{center}
 At long distances ($r\ge 2$ fm ) there exists weak attraction, which
is well understood and is dominated by the one pion exchange, while
contributions from the exchange of multi-pions and/or heavy mesons such as $\rho$, $\omega$ and $\sigma$ lead to slightly stronger attraction at medium distances (1 fm $\le r \le $ 2 fm).
On the other hand, at short distances ($r \le$ 1 fm), attraction turns into repulsion, and it becomes stronger and stronger as $r$ gets smaller, forming the strong repulsive core\cite{Jastrow}.
The repulsive core is essential not only for describing the $NN$ scattering data, but also for the stability and saturation of atomic nuclei, for determining the maximum mass of neutron stars, and for igniting Type II supernova explosions\cite{Supernova}.
Although the origin of the repulsive core must be related to the quark-gluon structure of nucleons,
it remains one of the most fundamental problems in nuclear physics for a long time\cite{OSY}.
It is a great challenge for us to derive the nuclear potential including the repulsive core from (lattice) QCD.

In this talk, we first explain  our strategy to extract $NN$ potentials theoretically from the first principle using lattice QCD and present the recent result in quenched QCD simulations. We then apply the method to various cases including the energy dependence of the potential, the quark mass dependence of the potential, the tensor potential, the full QCD calculation and the hyperon-nucleon potential. Since we mainly present only the results due to the limitation of the length, please see the corresponding references for more details.   

\section{Strategy to extract potentials in QCD}
Since a potential is a concept of the non-relativistic quantum mechanics, it is non-trivila to define it in QCD.  To find a reasonable definition of the $NN$ potentials in QCD, we first consider the $S$-matrix below inelastic threshold of the $NN$ scattering. The unitarity  leads to 
\begin{eqnarray}
S &=& e^{2 i \delta}
\end{eqnarray}
where the "phase" $\delta$ is a hermitian matrix.  We next introduce the equal time Bethe-Salpeter (BS) wave function\cite{BNNPEW,cppacs}, defined by
\begin{eqnarray}
\varphi_E({\bf r}) &=&\langle 0 \vert N ({\bf x} +{\bf r},0) N({\bf x},0) \vert 2N, E \rangle
\end{eqnarray}
where $\vert 2N, E\rangle$ is a two-nucleon eigenstate in QCD with energy $E=2\sqrt{m_N^2+k^2}$, and
the $N(x)$ is the gauge-invariant 3-quark operator given by
\begin{eqnarray}
N(x) &=& \varepsilon^{abc} q_a(x) q_b(x) q_c(x) .
\end{eqnarray}
For large $r=\vert {\bf r}\vert$, the partial wave $l$ of the BS-wave function behaves as
\begin{eqnarray}
\varphi_E^l({\bf r}) &\rightarrow & A_l \frac{\sin(kr-l\pi/2 +\delta_l(k))}{k r}
\end{eqnarray}
where $\delta_l(k)$ is the phase of the $S$-matrix for the partial wave $l$\cite{ishizuka,AHI2}. (Although $\delta_l(k)$ is a hermitian matrix in general, we here consider the case that $\delta_l(k)$ is just a number for simplicity.) The above formula says that $\delta_l(k)$ is the scattering phase shift of the
scattering wave. In other words, the BS wave function defined above can be interpreted as the $NN$ scattering wave.  

Based on the above fact, we have proposed the following strategy to define and extract the $NN$ potential in QCD\cite{IAH1,aoki}. We define a non-local potential from the BS wave function as
\begin{eqnarray}
\left[E-H_0\right] \varphi_E({\bf r}) &=& \int d^3 s\, U({\bf r}, {\bf s} )\varphi_E({\bf s}) 
\label{eq:def_pot}
\end{eqnarray}
where $H_0 = -{\bf \nabla}^2/(2\mu_N)$ and $\mu_N = m_N/2$ is the reduced mass of a two-nucleon system. 
Since the non-local potential is difficult to deal with, we expand it in terms of derivatives as
$U({\bf r},{\bf s}) = V({\bf r},{\bf \nabla})\delta^3({\bf r}-{\bf s})$. The first few terms
are given by\cite{OM}
\begin{eqnarray}
V({\bf r}, {\bf \nabla}) &=& V_C(r) + V_T(r) S_{12} + V_{\rm LS}{\bf \rm L}\cdot {\bf \rm S}
\nonumber \\
&+& \{ V_D(r), {\bf \nabla}^2 \} + \cdots , \\
S_{12} &=& \frac{3}{r^2} ({\bf \sigma_1}\cdot{\bf r}) ({\bf \sigma_2}\cdot{\bf r})
-({\bf \sigma_1}\cdot {\bf \sigma_2} )
\end{eqnarray}
where $S_{12} $ is the tensor operator  and $\sigma_i$ is the spin operator of
the $i$-th nucleon. The central potential $V_C$ and the tensor potential $V_T$ are the leading local terms (without derivatives), and thus can be determined from the BS wave function at one energy using
eq.(\ref{eq:def_pot}). Higer order terms in the above derivative expansion can be successively determined from BS wave functions at different energy. We however expect at low energy that  the leading order terms, $V_C$ and $V_T$, give a good approximation of the potential.
We can estimate an applicable range of energy for the local potential approximation, by
calculating physical observables such as the scattering phase shift  from the local potential and comparing them with experimental values.

The first result for the $NN$ potentials based on the above strategy has been obtained in quenched lattice QCD simulations\cite{IAH1}, where the lattice spacing $a$ is  0.137 fm, the spatial  extension $L$ is 4.4 fm  and the pion mass $m_\pi$ is 529 MeV. In Fig.\ref{fig:quench_potential}, the central potential for the $^1S_0$ (spin singlet and $L=0$) state and the effective central potential for the $^3S_1$ (spin triplet and $L=0$) state, obtained at $k^2 \simeq 0$, are plotted as a function of $r$.
By comparing Fig.\ref{fig:quench_potential} with Fig.\ref{fig:potential}, we see that qualitative features of $NN$ potentials are reproduced. Ref.\cite{IAH1} has been selected as one of 21 papers in Nature Research Highlights 2007\cite{nature}. 
\begin{center}
\includegraphics[width=60mm, angle=270]{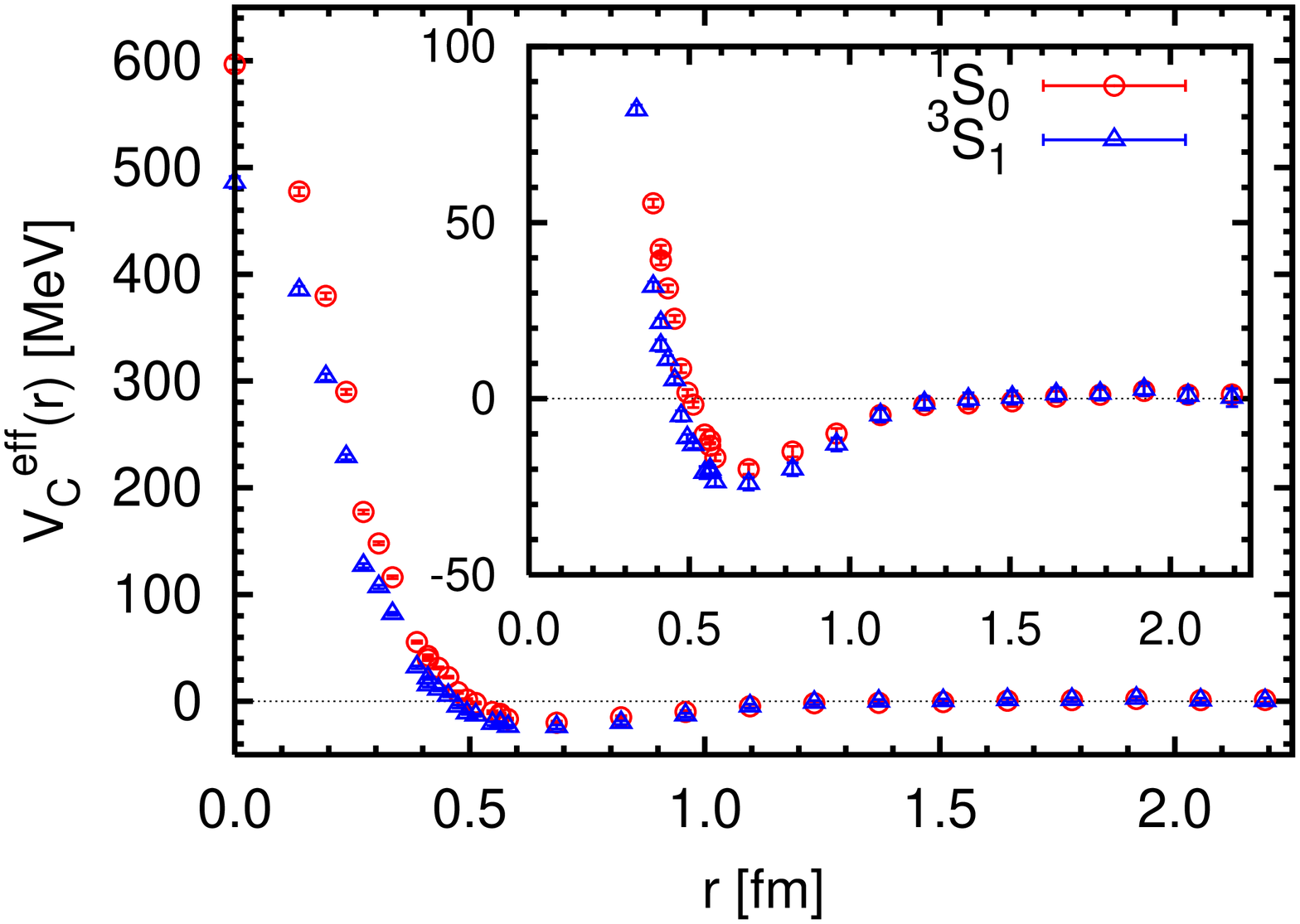}
\figcaption{\label{fig:quench_potential}
The (effective) central potential for the $^1S_0$ ($^3S_1$) state at $m_\pi = 529$ MeV in quenched QCD.
}
\end{center}

\section{Recent developments}
In this section we report recent developments of lattice QCD calculations for baryon-baryon interactions,
based on the method in the previous section.

\subsection{Energy dependence}
We first investigate the applicable range of energy for the local potential
determined at $k\simeq 0$.
If terms with derivatives such as $V_{\rm LS}(r) {\bf L}\cdot{\bf S}$ or $\{V_D(r), {\bf\nabla}^2\}$ becomes important, the local potential determined at $k > 0$ is different from the one at $k\simeq 0$.
From such $k$ dependences of  local potentials, in principle, 
some of the terms with derivatives can be determined. In Fig.\ref{fig:energy_dep}, the local potential for the $^1S_0$ state obtained at $k\simeq 250$ MeV (red, APBC) is compared with the one at $k\simeq 0$ MeV (blue, PBC) in quenched QCD at $a=0.137$ fm and $m_\pi = 529$ MeV\cite{ABHIMNW,murano}. 
\begin{center}
\includegraphics[width=82mm]{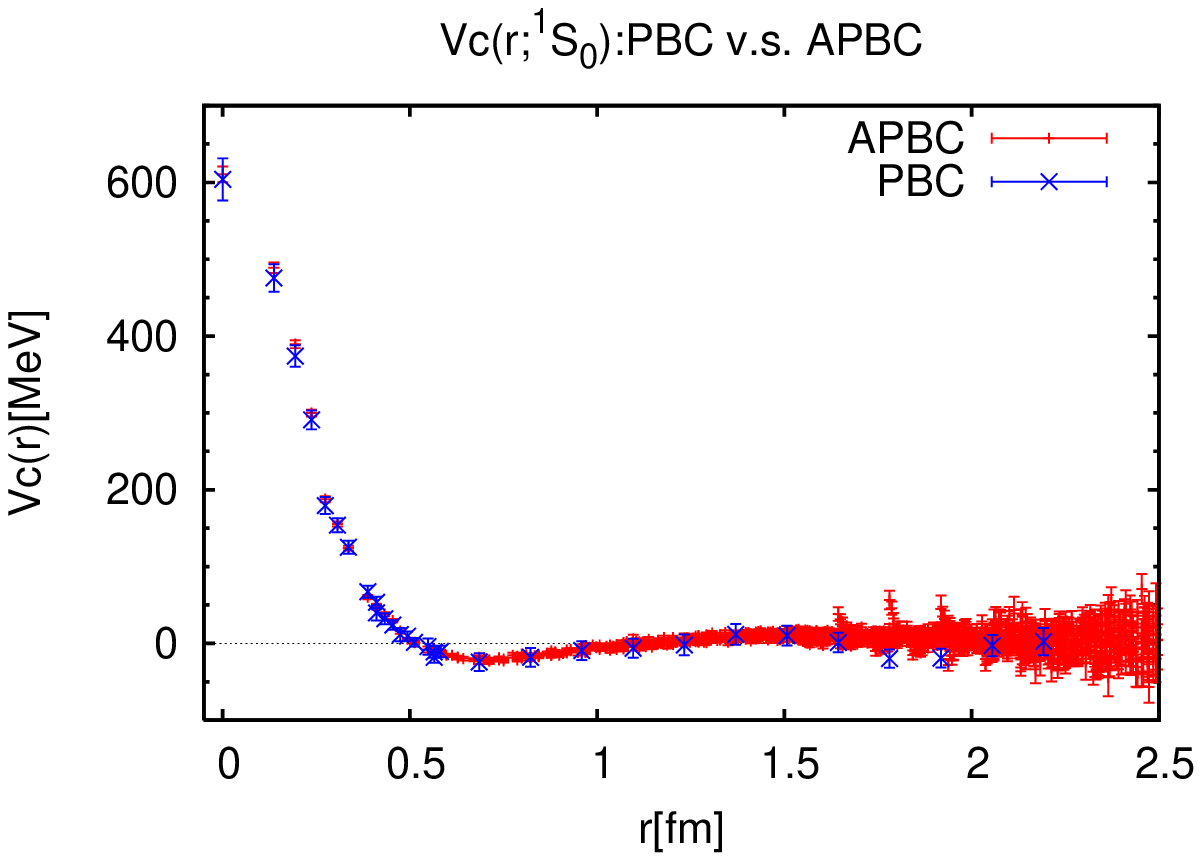}
\figcaption{\label{fig:energy_dep}
Comparison of central potentials in the local approximation between the periodic boundary condition (PBC, blue) and the anti-periodic boundary condition(APBC, red) for
the $^1S_0$ state at $m_\pi = 529$ MeV.}
\end{center}

As can be seen from the figure, the $k$ dependence of the local potential turns out to be very small.
This means that the potential obtained at $k\simeq 0$ in Fig.\ref{fig:quench_potential} well describes physical observables such as the phase shift $\delta_0(k)$  from $k\simeq 0$ to $k\simeq 250$ MeV, in quenched QCD at $a= 0.137$ fm and $m_\pi = 529$ MeV.

\subsection{Quark mass dependence}
A quark mass dependence of the $NN$ potential is shown in Fig.\ref{fig:mass_dep},
where the central potentials for the $^1S_0$ state
obtained at $k\simeq 0$ at $m_\pi = 380$, 529 and 731 MeV are compared
in quenched QCD  at $a=0.137$ fm\cite{AHI1,AHI2}. 

As quark mass decreases, the repulsion at short distance (the repulsive core) get stronger while
the attraction at intermediate distance ( 0.6 fm $\sim$ 1.2 fm) becomes also stronger.
\begin{center}
\includegraphics[width=58mm, angle=270]{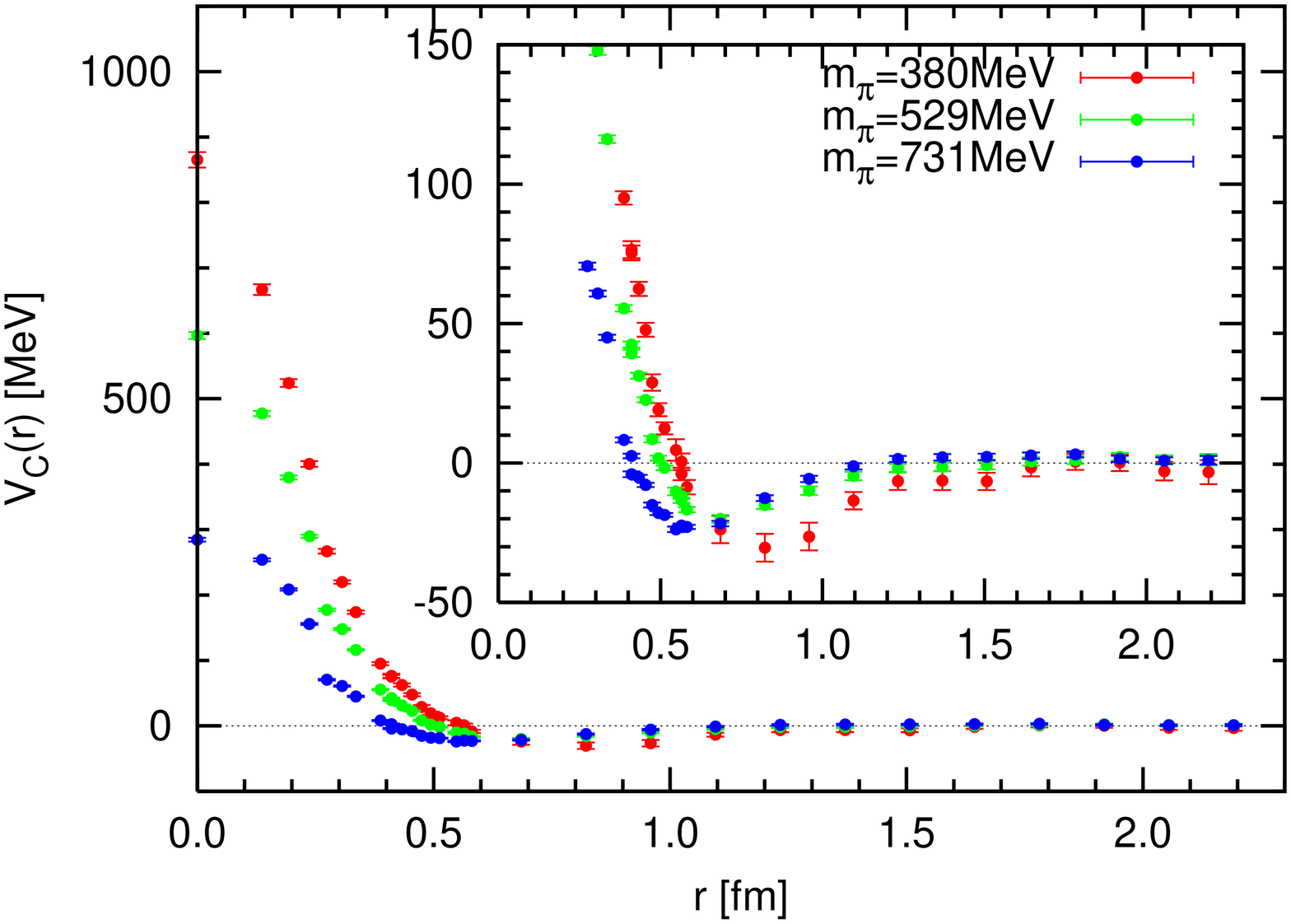}
\figcaption{\label{fig:mass_dep}
The central potential for  the $^1S_0$ state at $m_\pi = 380$ MeV (red),
529 MeV (green) and 731 MeV (blue) in quenched QCD at $a=0.137$ fm.}
\end{center}
 
\subsection{Tensor potential}
The tensor operator $S_{12}$ mixes the $^3S_1$( $J=S=1$ and $L=0$) state with the $^3D_1$ ($ J=S=1$ and $L=0$) state. Using this property, we can determine the tensor potential as follows.
For the $J=S=1$ state, the local potential approximation leads to
\begin{eqnarray}
(E-H_0)\varphi_E({\bf r} ) &=& \left[V_C(r) + V_T(r) S_{12}\right] \varphi_E({\bf r}),
\end{eqnarray}
which, by the projection $P$ to the $L=0$ state and the projection $Q$ to the $L=2$ state, is decomposed into
\begin{eqnarray}
\left(\begin{array}{cc}
P \varphi_E & P S_{12} \varphi_E \\
Q \varphi_E & Q S_{12} \varphi_E \\
\end{array}
\right) &\times &
\left(\begin{array}{c}
V_C \\
V_T \\
\end{array}
\right) \nonumber \\
&=& (E - H_0)
\left(\begin{array}{c}
P  \varphi_E\\
Q \varphi_E \\
\end{array}
\right) .
\end{eqnarray}
The above equation can be easily solved as
\begin{eqnarray}
\left(\begin{array}{c}
V_C \\
V_T \\
\end{array}
\right) &=&
\left(\begin{array}{cc}
P \varphi_E & P S_{12} \varphi_E \\
Q \varphi_E & Q S_{12} \varphi_E \\
\end{array}
\right)^{-1} \nonumber \\
&\times& (E - H_0)
\left(\begin{array}{c}
P  \varphi_E\\
Q \varphi_E \\
\end{array}
\right) .
\end{eqnarray}
In Fig.\ref{fig:tensor}, the central potential $V_C(r)$ and the tensor potential $V_T(r)$ for the
spin-triplet state are plotted\cite{AHI2}, together with the effective central potential $V_C^{\rm eff}(r)$ for the $^3S_1$ in Fig.\ref{fig:quench_potential}, which corresponds to $ (E-H_0) P \varphi_E / (P \varphi_E )$ in the above notation.
These potentials are calculated in quenched QCD at $a=0.137$ fm and $m_\pi=529$ MeV. 
\begin{center}
\includegraphics[width=60mm, angle=270]{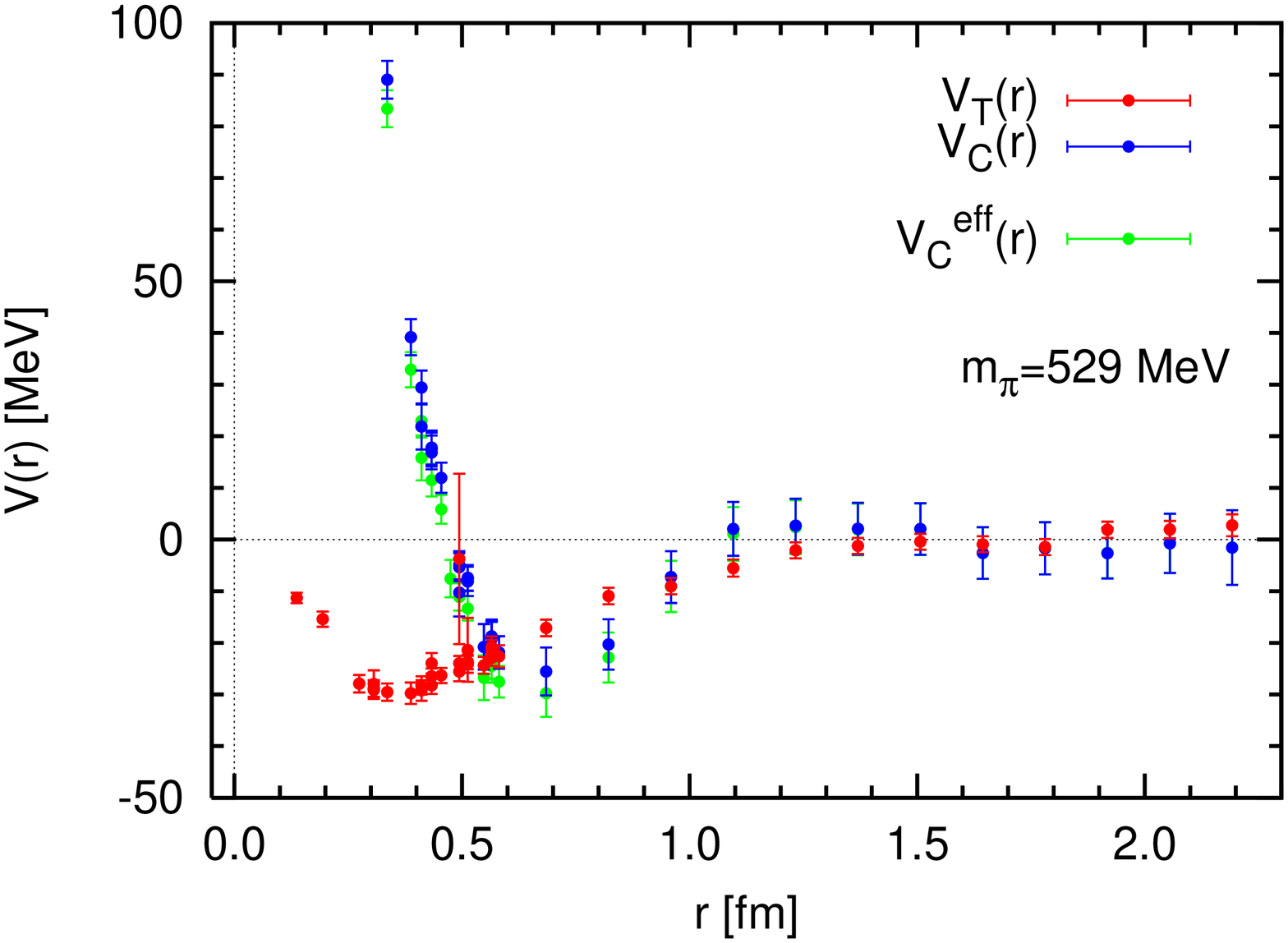}
\vspace{0.3cm}
\figcaption{\label{fig:tensor}
The central potential(blue) and the tensor potential (red), together with the effective central potential (red), for the spin-triplet state at $m_\pi =529$ MeV.
They are obtained from $k\simeq 0$ states in quenched QCD at $a=0.137$ fm.
}
\end{center}

We first notice that the tensor potential $V_T$ has no strong repulsive core, in contrast to the central potential. This feature is consistent with the previous phenomenological estimate\cite{Machleidt2}. 
Although the tensor potential is comparable in the magnitude with the central potential, the difference between the central and effective central potentials, which is caused by the second order perturbation of $V_T$, is very small at this quark mass.
A quark mass dependence of the tensor potential is given in Fig.\ref{fig:tensor_mass}, where
the tensor potential is plotted at $m_\pi = 380, 529$ and 731 MeV\cite{AHI2}.
The tensor potential becomes stronger as the quark mass decreases.
\begin{center}
\includegraphics[width=60mm, angle=270]{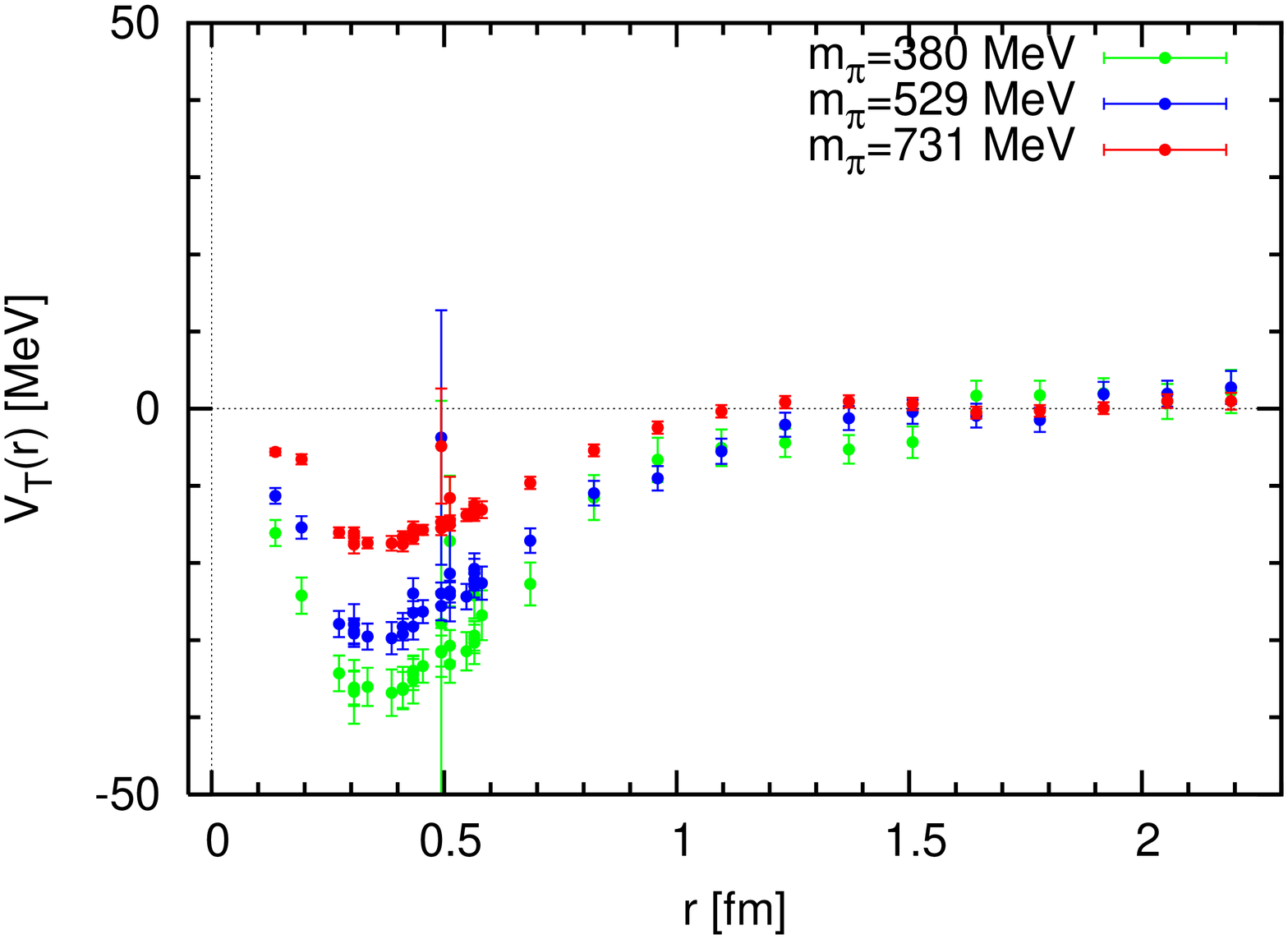}
\figcaption{\label{fig:tensor_mass}
Quark mass dependence of the tensor potential. }
\end{center}

\subsection{Full QCD calculations}
Results for the $NN$ potentials so far have been calculated in quenched QCD.
In Ref.\cite{IAH2}, preliminary results in full QCD calculations have been reported, based on 
gauge configurations generated by the PACS-CS Collaboration in 2+1 flavor QCD
at $a=0.09$ fm and $L=2.9$ fm\cite{pacs-cs}. Hadron spectra obtained from these configurations are shown in Fig.\ref{fig:spectra}. Agreements between lattice QCD predictions and experimental values are quiet good. 
The (effective) central $NN$ potentials on these configurations are given in Fig.\ref{fig:full_potential} for the $^1S_0$ state (red) and the $^3S_1$ state (blue) at $m_\pi = 702$ MeV.

\vspace{1.6cm}

\begin{center}
\includegraphics[width=80mm]{Fig/spectrum.eps}
\figcaption{\label{fig:spectra}
Light hadron spectra extrapolated to the physical point 
using $m_\pi$, $m_K$ and $m_\Omega$ as input. 
Horizontal bars denote the experimental values.
}
\end{center}
\begin{center}
\includegraphics[width=60mm,angle=270]{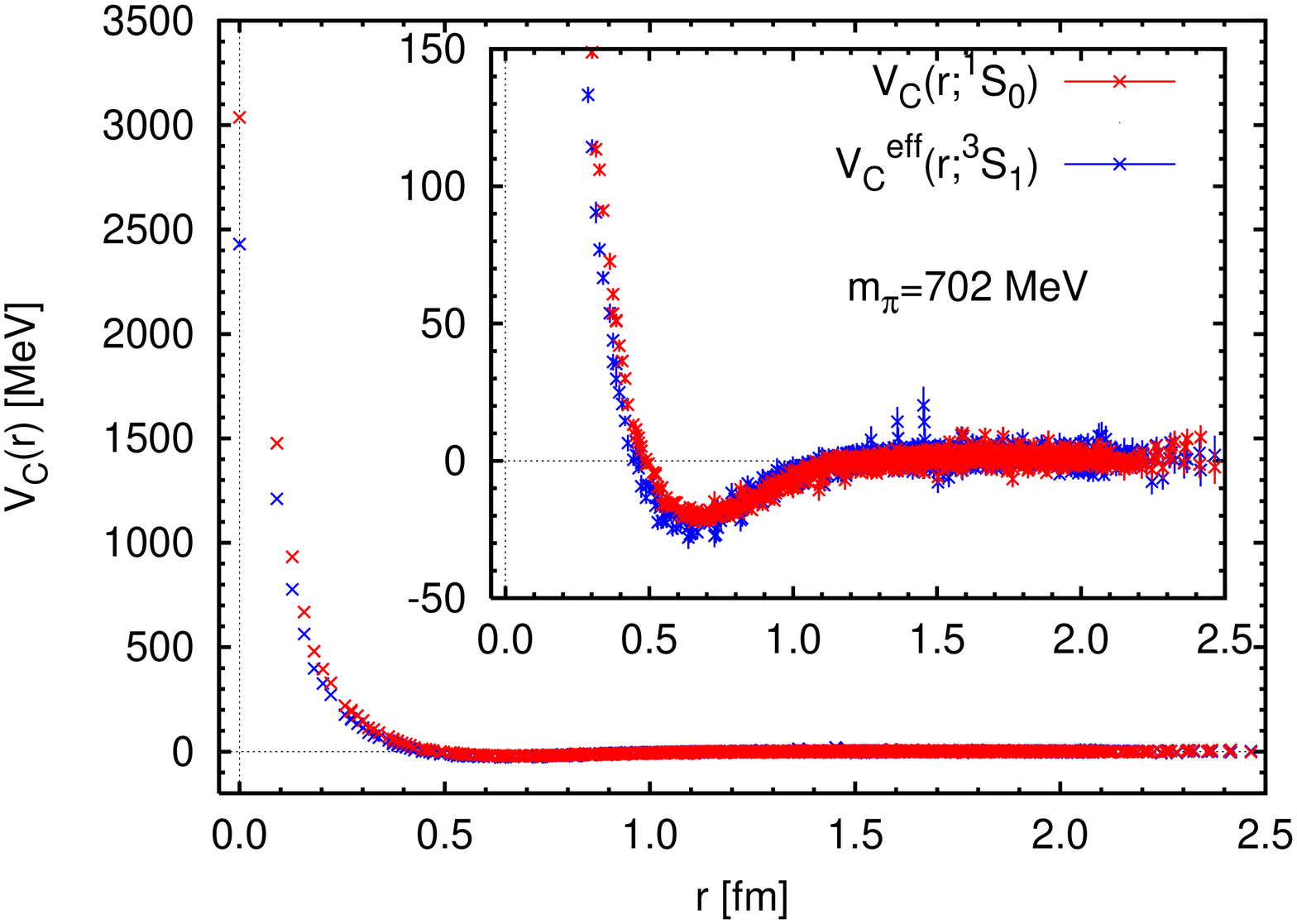}
\figcaption{\label{fig:full_potential}
The (effective) central potential for the $^1S_0$ state (red) and the $^3S_1$ state (blue) at $m_\pi = 702$ MeV
in 2+1 flavor QCD at $a=0.09$ fm.
}
\end{center}

We observe that the repulsive core in both states is much larger in magnitude than the corresponding one in quenched QCD in Fig.\ref{fig:quench_potential}, though the lattice spacing ( 0.09 fm vs. 0.137 fm) and the pion mass ( 529 MeV vs. 702 MeV ) are different.  A reason for the difference of the repulsive core between full and quenched QCD is now under investigation. 

\subsection{Hyperon-Nucleon interactions}
A hyperon is a baryon which contains at least one strange quark.
Contrary to the case of $NN$ interactions,
hyperon-nucleon ($YN$) or hyperon-hyperon($YY$) interactions can not be precisely determined, since the scattering experiments are either difficult or impossible 
due to the short life times of hyperons.
Our approach therefore may open a new possibility to determine them theoretically from QCD.
In this direction, potentials between a $\Xi^0$ (hyperon with strangeness $-2$ ) and a proton
has already been calculated in quenched QCD\cite{nemura1}.  
\begin{center}
\includegraphics[width=85mm]{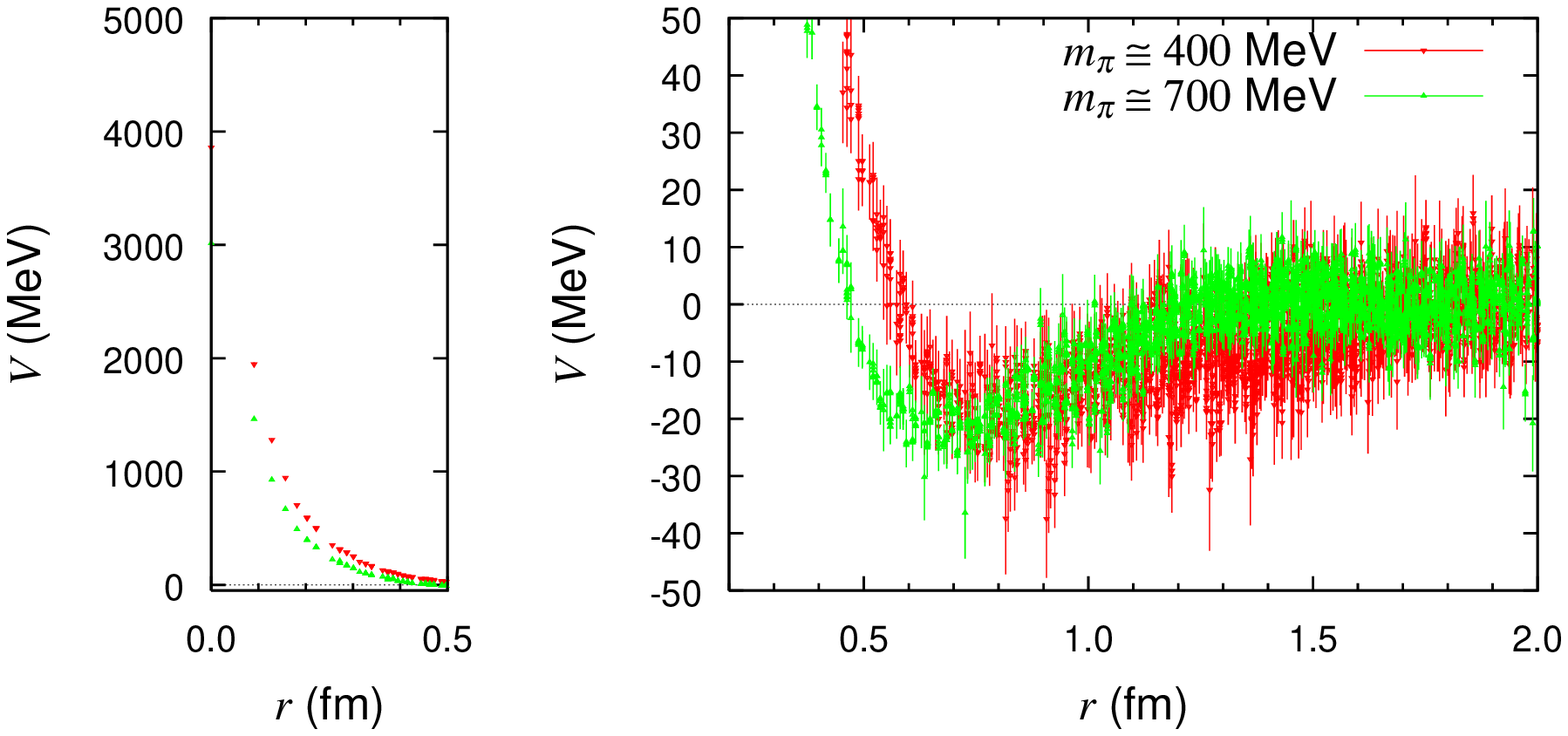}
\figcaption{\label{fig:hyperon_singlet}
The central potential for the $\Lambda N$ ($^1S_0$) state in 2+1 full QCD as a function of $r$ at $m_\pi \simeq 400$ MeV (red) and 700 MeV (green).
}
\end{center}
\begin{center}
\includegraphics[width=85mm]{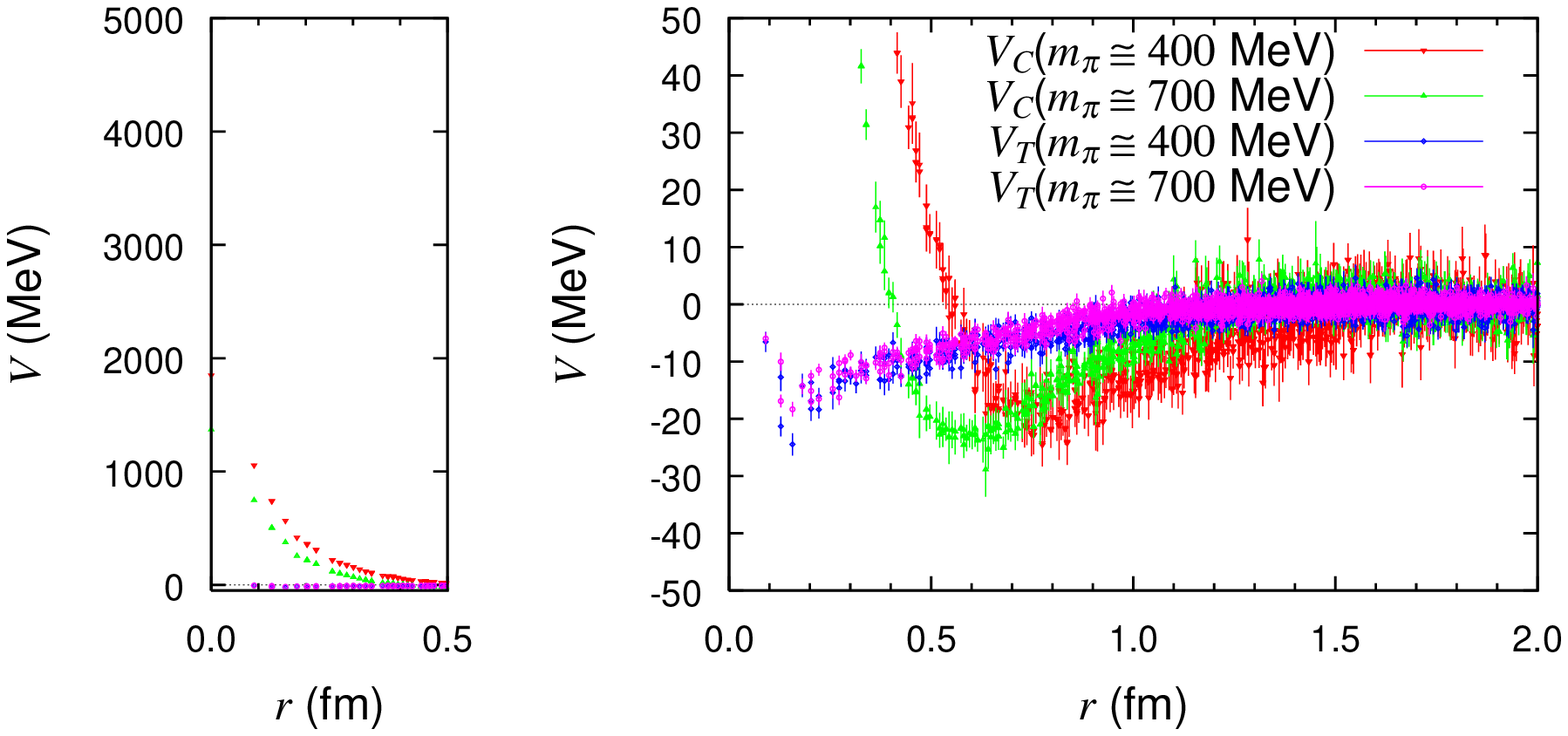}
\figcaption{\label{fig:hyperon_triplet}
The central and tensor potentials for the $\Lambda N$ ($^3S_1 - ^3D_1$) state in 2+1 full QCD at $m_\pi \simeq 400$ MeV (red and blue) and 700 MeV (green and magenta).
}
\end{center}

Recently calculations have been extended to the potentials between 
 the $\Lambda$( hyperon with strangeness $-1$ ) and $N$ in both quenched and full QCD. 
Fig.\ref{fig:hyperon_singlet} and Fig.\ref{fig:hyperon_triplet} show the $\Lambda N$ potentials
as a function of $r$ obtained from 2+1 flavor QCD calculation\cite{nemura2}  based on the PACS-CS gauge configurations.
The central  potential for the $^1S_0$ state  is given in Fig.\ref{fig:hyperon_singlet}, while the central and the tensor potentials for the  $^3S_1 - ^3D_1$ state are given in Fig.\ref{fig:hyperon_triplet} ,
with highlighting the short distance (medium to long distance) region in the left (right) panel.
These figures contain results  at $m_\pi \simeq 400$ and 700 MeV. 

As can be seen from both figures, the attractive well of the central potential moves to outer region as the quark mass decreases while the depth of these attractive pockets do not change so much.
The present results  show that the tensor force is weaker while the spin dependence is stronger than the $NN$ case\cite{IAH2}.  As in the $NN$ case, the hight of the repulsive core
is much larger than the quenched case\cite{nemura2} and it increases as the quark mass decreases.

\section{Conclusion}
We present recent results on baron-baryon interactions obtained from lattice QCD simulations.
In our strategy, baryon-baryon($NN$, $YN$ and $YY$) potentials are extracted from the BS wave functions.  The first result for the $NN$ (effective) central potentials in quenched QCD shows good "shape":  Qualitative features of the $NN$ potential have been reproduced in quenched QCD, and the energy dependence of the potentials is weak at low energy.
The method has been successfully extended to the tensor potential and the $\Lambda N$ potentials in both quenched and full QCD.

One of the ultimate goal in our approach is to calculate baryon-baryon potentials in full QCD at $m_\pi = 140$ MeV. In such calculations one can investigate, for example, a relation between the deuteron binding and the tensor force.  As other directions of our approach, it is important to extract the 3-body force\cite{miyazawa} from QCD and  to understand the origin of the repulsive core theoretically\cite{ABW}.
 
\acknowledgments{I would like to thank members of HAL QCD Collaboration for providing me the latest results and useful discussions. This work is supported in part by 
Grants-in-Aid of the Ministry of Education, Culture, Sports, Science and Technology of Japan (Nos. 20340047, 20105001, 20105003). }

\end{multicols}

\vspace{-2mm}
\centerline{\rule{80mm}{0.1pt}}
\vspace{2mm}

\begin{multicols}{2}

\end{multicols}
\clearpage

\end{document}